\documentclass[prl,aps,showpacs,twocolumn,floatfix,amsmath,amssymb]{revtex4}
\usepackage{graphicx}

\def \be#1\ee {\begin{equation}#1\end{equation}}
\def \bem#1\eem {\begin{multline}#1\end{multline}}
\def \beg#1\eeg {\begin{gather}#1\end{gather}}
\def \bea#1\eea {\begin{align}#1\end{align}}

\newcommand{\eps}{\varepsilon}
\newcommand{\tQ}{\tilde Q}
\newcommand{\tQs}{\tilde{\underline{Q}}\vphantom{Q}}
\newcommand{\tEg}{\tilde E_g}
\newcommand{\Etun}{E_g^{\mathrm{tun}}}

\newcommand{\calQ}{\mathcal{Q}}

\newcommand{\tx}{\hat\tau_1}
\newcommand{\ty}{\hat\tau_2}
\newcommand{\tz}{\hat\tau_3}

\newcommand{\const}{\text{const}}

\DeclareMathOperator{\Tr}{Tr}
\DeclareMathOperator{\tr}{tr}

\begin{document}

\title{Coulomb Blockade of Proximity Effect at Large Conductance}

\author{P.M.Ostrovsky}
\email{ostrov@itp.ac.ru}
\affiliation{Landau Institute for Theoretical Physics, Chernogolovka, Moscow
region, 142432 Russia}

\author{M.A.Skvortsov}
\affiliation{Landau Institute for Theoretical Physics, Chernogolovka, Moscow
region, 142432 Russia}

\author{M.V.Feigel'man}
\affiliation{Landau Institute for Theoretical Physics, Chernogolovka, Moscow
region, 142432 Russia}
\affiliation{Materials Science Division, Argonne National Laboratory,
 Argonne, Illinois 60439, USA}

\begin{abstract}
We consider the proximity effect in a normal dot coupled to a bulk
superconducting reservoir by the tunnel contact with large normal conductance.
Coulomb interaction in the dot suppresses the proximity minigap induced in the
normal part of the system. We find exact expressions for the thermodynamic and
tunneling minigaps as functions of the junction's capacitance. The tunneling
minigap interpolates between its proximity-induced value in the regime of weak
Coulomb interaction to the Coulomb gap in the regime of strong interaction. In
the intermediate case a non-universal two-step structure of the tunneling
density of states is predicted. The charge quantization in the dot is also
studied.
\end{abstract}


\maketitle

When a normal metal forms a contact with a superconductor it acquires some
superconductor features. One of them is the suppression of the electron density
of states (DoS) at the Fermi energy. This effect is governed by the Andreev
processes at the normal-metal--superconductor (NS) boundary. Cooper pairs
tunnel into the normal metal preserving their phase and thus inducing the
superconductive correlations. Coulomb repulsion leads to phase fluctuations and
reduces this proximity effect~\cite{Fazio94}. For a disordered normal metal, a
minigap of the order of the inverse escape time appears in the excitation
spectrum~\cite{Melsen97}. Below we study the suppression of the minigap by
Coulomb interaction.

To measure the density of states the tunneling spectroscopy technique is widely
employed. The conductance as a function of the voltage between the external
probe and the system under investigation is proportional to the
\emph{tunneling} density of states (TDoS). Coulomb interaction affects the TDoS
suppressing tunneling conductance at small voltages. This suppression is known
as the zero bias anomaly~\cite{AltshulerAronov79,LevitovShytov97}. The
interplay between the zero bias anomaly and the proximity effect was first
studied in Ref.~\cite{Oreg99} for a 2D thin normal film coupled by the tunnel
junction to a superconductor. The renormalization group procedure  yields the
power-law suppression of the minigap in the TDoS by the Coulomb interaction.

Coulomb repulsion is ultimately strong in restricted geometry. In such systems
direct observation of charge quantization is made possible by the Coulomb
blockade effect~\cite{GrabertDevoret92}. When a zero-dimensional (0D) normal
grain is coupled to a \emph{normal} reservoir by the tunnel junction with small
dimensionless [in units of $e^2/\hbar$] conductance $G$, the equilibrium charge
of the grain is a step function of the gate voltage at zero temperature.
Contrary, if $G\gg1$ the charge of the grain is no longer conserved and only
exponentially weak modulation of charge-voltage dependence
remains~\cite{WangGrabert96}. The situation changes dramatically if the
reservoir is \emph{superconducting}~\cite{MatveevGlazman98}. Now a single
electron cannot escape into the superconductor and charge quantization is
observed even at large normal conductance.

In this Letter we consider the 0D NS system described above. The parameters of
the system are assumed to satisfy the following conditions:
\be
  E_{\mathrm{Th}} > \Delta \gg (E_g, E_C) \gg \delta.
\label{params}
\ee
Here $E_{\mathrm{Th}}$ is the Thouless energy, $\Delta$ is the superconductor
gap, $\delta$ is the electron mean level spacing in the grain per one spin
component, $E_g = G\delta/4$ is the proximity minigap~\cite{Melsen97} in the
absence of Coulomb repulsion, and $E_C = e^2/(2C)$ is the charging energy, with
$C$ being the junction's capacitance. The dimensionless conductance $G$ is
assumed to be large. We neglect effects of quasiparticle transport assuming
temperature sufficiently low and thus quasiparticle conductance
$G_{\mathrm{qp}} = G\exp(-\Delta/T) \ll 1$.

Coulomb interaction modifies the thermodynamic and tunneling DoS in different
ways. The minigap $\tEg$ in the thermodynamic DoS is gradually suppressed with
the increase of $E_C$ due to enhanced phase fluctuations, whereas the
dependence of the minigap $\Etun$ in the TDoS on $E_C$ is more complicated.
Being suppressed at weak interaction by phase fluctuations, it eventually
reaches the Coulomb gap $E_C$ in the strong interaction limit. Qualitatively,
these two regimes are distinguished by the relation between the charging energy
$E_C$ and the energy $E_J = (E_g^2/\delta)\log(\Delta/E_g)$, which is the
Josephson energy of the fictitious system where the normal grain is replaced by
the weak superconductor with the gap $E_g$. Below we find the exact dependence
of both the DoS and TDoS on the strength of the interaction for arbitrary
$E_C/E_J$.

We also study the charge quantization in the grain accounting for both Coulomb
and proximity effects. The result of Ref.~\cite{MatveevGlazman98} is valid only
for $E_C \gg E_J$. In the opposite case proximity coupling smears the Coulomb
staircase preserving however the sharp discontinuities at half-integer charge.

To investigate the electron properties of the NS system we use the non-linear
$\sigma$-model~\cite{Finkelstein90} in the Matsubara representation. Disorder
average is made with the help of replica trick. Thus the $\sigma$-model is
formulated in terms of matrix field $Q$ operating in the Nambu-Gor'kov (Pauli
matrices $\hat\tau_i$), energy, and replica spaces. The 0D Coulomb interaction
is decoupled by the electric potential $\phi$. The $\sigma$-model action
reads~\cite{Finkelstein90}
\be
 S[Q,\phi]
  =
    -\frac{\pi}{\delta}\Tr \left[
      (\eps\tz+\phi)Q
    \right]
    -\frac{\pi G}4 \Tr[Q_SQ]
    + \int d\tau\frac{\phi^2}{4E_C}.
 \label{actionQphi}
\ee
Hereafter we omit replica indices for brevity. We will use $\tx$ for the
$Q$-matrix of the superconductor, $Q_S$, restricting our model to the subgap
region $\eps\ll\Delta$. The contribution from higher energies
leads~\cite{LarkinOvchinnikov83} to the renormalization of the capacitance $C
\mapsto C +e^2G/(2\Delta)$ and corresponding renormalization of the charging
energy $E_C$.

In the action~(\ref{actionQphi}) the matrix $Q$ is linearly coupled to the
potential $\phi$. This means that $Q$ contains not only soft electron modes of
the system with energies close to the Fermi energy but also the high energy
fluctuations corresponding to the shift of the whole energy band by electric
potential. To get rid of this contribution we make a gauge transformation
proposed in Ref.~\cite{KamenevAndreev99}: $Q_{\tau\tau'} = e^{i\tz K(\tau)}
\tQ_{\tau\tau'} e^{-i\tz K(\tau')}$
with $K(\tau) = \int^{\tau}\phi(\tau)d\tau$.
In terms of these new variables the action has the form
\bem
 S[\tQ,K]
  = -\frac{\pi}{\delta}\Tr(\eps\tz\tQ)\\
    +\int d\tau \left[
      \frac{\dot K^2}{4E_C}
      -\frac{2\pi E_g}{\delta} \left(
        \tQ^{(1)}_{\tau\tau}\cos 2K
        +\tQ^{(2)}_{\tau\tau}\sin 2K
      \right)
    \right],
 \label{action}
\eem
where the symbols $\tQ^{(i)} = \tr(\hat\tau_i\tQ)/2$ denote the Nambu-Gor'kov
components of the matrix $\tQ$.

The general dynamics governed by the action~(\ref{action}) is complicated as
the variables $\tQ$ and $K$ are strongly coupled with each other. Fortunately,
in the region of the parameters specified by Eq.~(\ref{params}) the
characteristic frequencies of the variable $\tQ$ (of the order of the
renormalized proximity gap $\tEg$) appear to be much smaller than those of the
variable $K$. Therefore, it is possible to employ the \emph{adiabatic}
approximation, integrating first over the ``fast'' variable $K$. The resulting
effective action $S[\tQ]$ for the ``slow'' variable $\tQ$ should then be
subject to the saddle-point approximation (SPA). Fluctuations of $Q$-matrix
describe the effects of level statistics, therefore the SPA works fine if the
relevant energy scale $\tEg$ exceeds $\delta$. Such a condition is definitely
satisfied in the regime of weak Coulomb repulsion: $E_g/\delta = G/4 \gg 1$.
Below we will find the applicability range both of the adiabatic and
saddle-point approximations at arbitrary strength of the Coulomb interaction.

The action averaged over $K$ is obviously uniform in time that allows to assume
the stationary saddle point
\be
 \tQs_{\eps\eps'}
  = 2\pi\delta(\eps-\eps')
    (\tz\cos\alpha(\eps) + \tx\sin\alpha(\eps)),
 \label{saddle_point}
\ee
parametrized by the single function $\alpha(\eps)$. Possibility to choose
$\tQs$ in the form (\ref{saddle_point}) without the $\ty$ term is related to
the zero mode with respect to $K\mapsto K+\const$.

Substituting the ansatz~(\ref{saddle_point}) into the action~(\ref{action})
we find the imaginary time evolution for $K$. It is determined by the simple
Hamiltonian
\be
 \hat H
  = E_C\left[
      -\partial^2/\partial K^2 - 2q\cos 2K
    \right].
 \label{hamilton}
\ee
The parameter $q = (E_g/2E_C\delta)\int_{-\Delta}^{\Delta}d\eps
\sin\alpha(\eps)$ measuring the relative strength of the proximity coupling
with respect to Coulomb interaction is to be determined self-consistently later
on. The $2\pi$-periodic boundary conditions for $K$ are implied.

At zero temperature the dynamics of $K(\tau)$ is frozen at the ground state
of~(\ref{hamilton}). The ground state energy can be expressed in terms of
zeroth Mathieu characteristic value $a_0(q)$: $E_0 = E_C a_0(q)$. After
averaging over $K$ one obtains the effective action for $\tQ$:
\be
 S[\tQ]
  = \int d\tau\left[
      -\frac{1}{\delta} \int d\eps\; \eps\cos\alpha(\eps)
       + E_0 (q)
    \right].
 \label{StQ}
\ee
Minimizing this expression we find the BCS-like solution $\alpha(\eps) =
\arctan(\tEg/\eps)$. The renormalized value of the minigap $\tEg$ satisfies the
following system of equations:
\be
  \frac{\tEg}{E_g}
   = -\frac{1}{2E_C}\frac{\partial E_0}{\partial q},
 \qquad
  q
   = \frac{E_g\tEg}{E_C\delta}\log\frac{2\Delta}{\tEg}.
 \label{selfconsistency}
\ee

The dependence of the ground state energy $E_0$ on $q$ is especially simple in
two limiting cases. If the Coulomb interaction is weak the parameter $q$ is
large and we can estimate the potential energy in Eq.~(\ref{hamilton}) by the
oscillator potential. This gives a small correction to the bare minigap value
$E_g$:
\be
 \tEg
  = E_g - \frac 12 \sqrt{\frac{E_C\delta}{\log(2\Delta/E_g)}},
 \qquad
 q \gg 1.
 \label{largeq}
\ee
In this regime the variable $K$ has the frequency $E_C\sqrt q$ which is
$\sqrt{E_C/\delta}$ times larger than $E_g$. In the opposite limit of strong
Coulomb interaction the potential in Eq.~(\ref{hamilton}) can be treated
perturbatively yielding $a_0(q) = -q^2/2$. The minigap appears to be
exponentially small in $E_C$:
\be
 \tEg
  = 2\Delta \exp\left(-\frac{2E_C\delta}{E_g^2}\right),
 \qquad
 q \ll 1.
 \label{smallq}
\ee
At the same time, the frequency of $K$ is $E_C \gg \tEg$,
ensuring the validity of the adiabatic approximation.

\begin{figure}
\includegraphics[width=0.95\columnwidth]{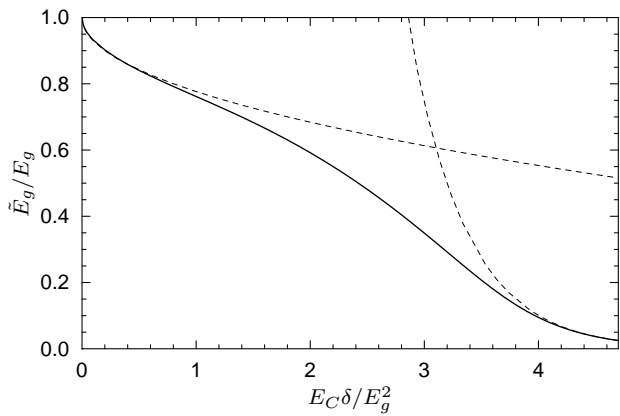}
\caption{The minigap in the thermodynamic DoS as a function
of the Coulomb energy $E_C$. Dashed lines illustrate the asymptotic
behaviors~(\ref{largeq}) and~(\ref{smallq}). The crossover from weak to strong
fluctuations regime occurs at $E_C\delta/E_g^2 \sim \log(\Delta/E_g)/2$. This
figure shows the case $\log(\Delta/E_g) = 5$.}
\label{Fig:tEg}
\end{figure}

The whole dependence of $\tEg$ on $E_C$ is shown in Fig.~\ref{Fig:tEg}. The
thermodynamic DoS itself is just the conventional BCS density of states with
the gap $\tEg$.

\begin{figure}
\includegraphics[width=0.95\columnwidth]{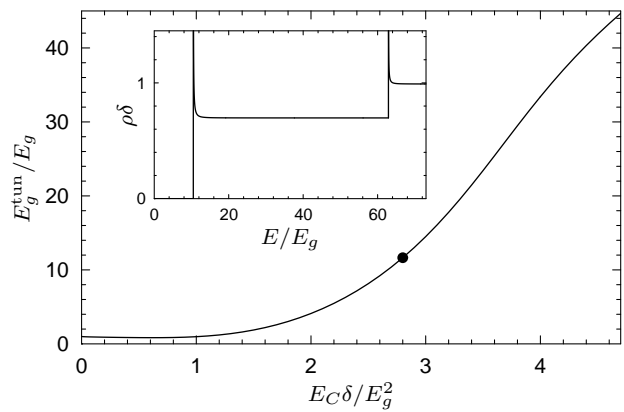}
\caption{The minigap in the tunneling DoS as a function of the Coulomb energy.
For small $E_C$ it follows $\tEg$ given by Eq.~(\ref{largeq}). In the opposite
limit $\Etun = E_C$. This figure is plotted assuming $\log(\Delta/E_g) = 5$ and
$G = 40$. Inset: two-step tunneling density of states in the intermediate
regime. The charging energy is such that $E_C\delta/E_g^2 = 2.8$ (as shown by
the dot on the main graph) which corresponds to $q = 0.95$.}
\label{Fig:Etun}
\end{figure}

Now we turn to the tunneling density of states. In the $\sigma$-model formalism
the TDoS is the analytic continuation of the function
\be
 \rho(\eps)
  = \frac 1{\delta} \int d\tau\, e^{i\eps\tau}\tr\left<
        \tz e^{i\tz K} \tQ e^{-i\tz K}
    \right>_{\tau,0}
\ee
from positive Matsubara energies $\eps$ to real energies $E$. The angular
brackets imply averaging with the action~(\ref{action}). In the adiabatic
approximation this means averaging over $K$ with the
Hamiltonian~(\ref{hamilton}) and substitution $\tQ \to \tQs$. The result for
the TDoS is the convolution of the thermodynamic DoS with the phase correlator
$C(\tau) = \left<e^{i(K(\tau)-K(0))}\right>$ given by
\be
 C(\omega)
  = \sum\limits_n
     \left|\langle 0|e^{iK}|n\rangle\right|^2
     \frac{2(E_n - E_0)}{(E_n-E_0)^2 + \omega^2}
 \label{Comega}
\ee
in the frequency representation at zero temperature. Here $E_i$ are the energy
levels of the Hamiltonian~(\ref{hamilton}). The matrix elements in
Eq.~(\ref{Comega}) are non-zero only for $n = 4k+1$ and $n = 4k+2$. Evaluating
the convolution of the thermodynamic DoS with $C(\omega)$ and performing
analytic continuation, we obtain for the TDoS:
\be
  \rho(E)
  =
  \frac{2}{\delta}
  \sum\limits_n
    \left|\langle 0|e^{iK}|n\rangle\right|^2
    \frac{|x_n|\,\theta(|x_n|-\tEg)}
         {\sqrt{x_n^2 - \tEg^2}} ,
\label{TDOS}
\ee
where $x_n=E-(E_n-E_0)$.
The tunneling minigap associated with the $n=1$ term is given by
\be
 \Etun
  = \tEg + E_1 - E_0.
 \label{Etun}
\ee
Fig.~\ref{Fig:Etun} shows its dependence on $E_C$ which
interpolates from the proximity gap (\ref{largeq}) in the weak
interaction regime to the Coulomb gap $E_C$ at large $E_C$. Note
the reentrant behavior of $\Etun$ as a function of $E_C$ at small
$E_C$.

The form of the TDoS depends on the strength of the Coulomb interaction. Due to
the fast decrease of the matrix element $\left|\langle
0|e^{iK}|n\rangle\right|^2$ with $n$ only the first two terms may significantly
contribute to the sum in Eq.~(\ref{TDOS}). In the weak Coulomb interaction
regime ($q \gg 1$), the $n=1$ contribution dominates. Since the level splitting
$E_1 - E_0$ is exponentially small the TDoS coincides with the thermodynamic
DoS in this limit. When the Coulomb interaction is strong ($q \ll 1$), the
matrix elements $\left|\langle 0|e^{iK}|1\rangle\right|^2 \approx \left|\langle
0|e^{iK}|2\rangle\right|^2 \approx 1/2$ and $\rho(E) \approx (2/\delta) \,
\theta(|E|-E_C)$. In the intermediate regime ($q \sim 1$) the TDoS acquires a
non-universal two-step structure, being zero for $E < \Etun$, of the order of
$\left|\langle 0|e^{iK}|1\rangle\right|^2$ for $\Etun < E < \tEg + E_2 - E_0$,
and close to 1 for $E > \tEg + E_2 - E_0$, see the inset of
Fig.~\ref{Fig:Etun}. The BCS-type singularities are still present at the step
edges but they are relatively weak.

At non-zero temperature all above results are still valid provided $T \ll
\tEg$. If the temperature is higher all integrations over energies should be
replaced by summations over Matsubara frequencies. Moreover, the phase $K$ is
not frozen at the ground state in this case, and the temperature dependent free
energy should be used instead of the ground state energy in Eq.~(\ref{StQ}) and
below. These calculations show that the true tunneling minigap vanishes when
the temperature is of the order of $\tEg(T=0)$.

Now we briefly comment on the validity of our approach. The above results were
obtained using the adiabatic approximation for the integral over $K$ and the
saddle point approximation for $\tQ$. To justify this approach one can
calculate the first fluctuation correction to the TDoS. To this end we
parametrize $\tQ$ in the standard way in terms of the matrix $W$, and average
over $K$ with the Hamiltonian~(\ref{hamilton}) taking into account $W$
perturbatively. As a result we end up with the quadratic action for the
fluctuating $W$. Evaluating the first order perturbative correction to the
density of states we find that it is small provided that $E_C\gg\delta$
(adiabaticity condition) and $\tEg \gg \delta$ (validity of the SPA). All the
above results were obtained in the lowest-order approximation over tunneling
probability ${\cal T}$ across NS interface. The effect of higher-order terms
can be neglected if Andreev conductance through this interface is small, $G_A
\ll 1$.

Finally, we consider the case of a non-zero gate voltage $V$ applied to the
normal grain in the standard Coulomb blockade setup~\cite{MatveevGlazman98}. If
the grain is coupled to the gate by the capacitance $C_g$ the charging
energy~\cite{GrabertDevoret92} becomes $E_C(\hat\calQ/e-N)^2$, where
$\hat\calQ$ is the charge operator of the grain, $eN=C_gV$ is the equilibrium
charge, and we redefined $E_C = e^2/2(C+C_g)$. The Hamiltonian for $K$ is
changed to
\be
 \hat H
  = E_C\left[
      \bigl(
        -i\partial/\partial K - N
      \bigr)^2 - 2q\cos 2K
    \right].
 \label{hamilton2}
\ee
The energy spectrum of this Hamiltonian has the standard band structure. The
boundary conditions pick out \emph{two} energy levels from each band with the
quasimomentum determined by the gate potential. When the equilibrium charge $N$
approaches half-integer values these two energy levels cross. Since all
physical quantities depend periodically on $N$, in what follows we assume $|N|
< 1/2$.

\begin{figure}
\includegraphics[width=0.95\columnwidth]{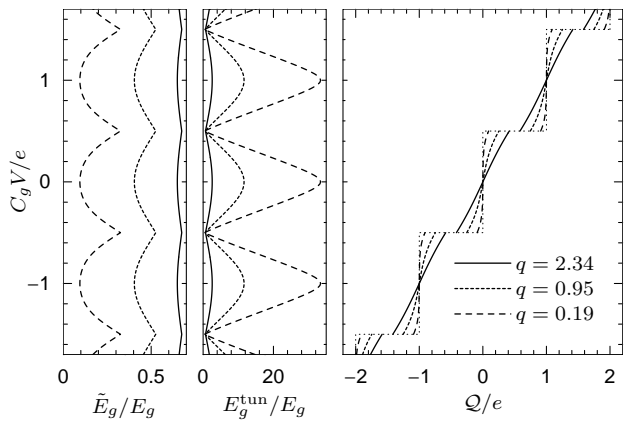}
\caption{The dependence of thermodynamic minigap, tunneling minigap, and average
charge on the gate voltage $V$ (vertical axis) at three different values of
Coulomb energy. The parameters are $G = 40$, $\log(\Delta/E_g) = 5$.}
\label{Fig:gate}
\end{figure}

In the weak Coulomb interaction regime, $E_C \ll E_J$, the lowest energy band
is exponentially narrow. The lowest level of the Hamiltonian depends weakly on
$N$, and all the results for the DoS and TDoS obtained above are left
unchanged. For strong Coulomb interaction we calculate the ground state energy
perturbatively in small $q$. Neglecting the $q$ term we have $E^{(0)}_n = E_C(n
- N)^2$. The second order in $q$ correction to the ground state energy
$E^{(2)}_0 = -E_C q^2/2(1 - N^2)$. From the self-consistency
equations~(\ref{selfconsistency}) we find the thermodynamic minigap $\tEg =
2\Delta\exp[-(2E_C\delta/E_g^2)(1-N^2)]$. The minigap in the TDoS is determined
by Eq.~(\ref{Etun}) yielding $\Etun = \tEg + E_C(1-2|N|)$. This quantity
depends strongly on $N$ decreasing to the exponentially small value $\tEg$ at
$N=1/2$. These results are illustrated in Fig.~\ref{Fig:gate}.

The average charge of the grain is $\calQ = eN - \partial E_0/\partial V$. In
the limit of strong Coulomb interaction the result of~\cite{MatveevGlazman98}
is reproduced. In the opposite limit the Coulomb staircase is smeared up to an
exponentially weak modulation. Employing the tight binding approximation for
the Hamiltonian~(\ref{hamilton2}) we get $\calQ/e = N -
C_g/(C+C_g)8\sqrt{2\pi}q^{3/4}e^{-4\sqrt q} \sin(\pi N)$. The small steps of
the height $C_g/(C+C_g)16\sqrt{2\pi}q^{3/4}e^{-4\sqrt q}$ at half-integer
charge are still present as predicted in Ref.~\cite{MatveevGlazman98}. This
feature is the consequence of the double-electron charge transport through the
junction. The Coulomb staircase is shown in the right part of
Fig.~\ref{Fig:gate}.

In conclusion, we have shown that interplay between proximity and charging
effects in a SIN structure with large normal conductance $G$ is governed by the
ratio of the charging energy $E_C$ to the fictitious Josepshson energy $E_J =
G^2\delta\log(\Delta/G\delta)$. DoS and TDoS are calculated as functions of
$E_C/E_J$ and the gate voltage $V$. An unsual two-step shape of the TDoS is
found at $E_C\sim E_J$. We are grateful to D. Bagrets and A. I. Larkin for
useful discussions. This research was supported by the Russian Ministry of
Science, Russian Academy of Sciences, and RFBR grant No.\ 01-02-17759; Landau
Scholarship, Juelich (P.M.O.), the Dynasty Foundation and the ICFPM (M.A.S. and
P.M.O.), and by the US DoE through contract No. W-31-109-ENG-38 (M.V.F.).

\end{document}